\documentclass[
amsmath,amssymb,
aps,
reprint,%
]{revtex4-1}

\usepackage{color}

\def\siml{\hspace{1ex} ^{<} \hspace{-2.5mm}_{\sim} \hspace{1ex}}

\usepackage{graphicx}
\usepackage{dcolumn}
\usepackage{bm}

\begin{document}


\title[Simple improvements to classical bubble nucleation  models]
{Simple improvements to classical bubble nucleation models 
 }

\author{Kyoko K. Tanaka}
\affiliation{Institute of Low Temperature Science, Hokkaido University, Sapporo 060-0819, Japan} 

\author{Hidekazu Tanaka}
\affiliation{Institute of Low Temperature Science, Hokkaido University, Sapporo 060-0819, Japan} 

\author{Raymond Ang\'elil} \affiliation{Institute for Computational
Science, University of Z\"urich, 8057 Z\"urich, Switzerland}

\author{J\"urg Diemand}
\affiliation{Institute for Computational Science, University of Z\"urich, 8057 Z\"urich, Switzerland}

\date{\today}

\begin{abstract}

  We revisit classical nucleation theory (CNT) for the homogeneous
  bubble nucleation rate and improve the classical formula using a new
  prefactor in the nucleation rate.  Most of the previous theoretical
  studies have used the constant prefactor determined by the bubble
  growth due to the evaporation process from the bubble
  surface. However, the growth of bubbles is also regulated by the
  thermal conduction, the viscosity, and the inertia of liquid
  motion. These effects can decrease the prefactor significantly,
  especially when the liquid pressure is much smaller than the
  equilibrium one. The deviation in the nucleation rate between the
  improved formula and the CNT can be as large as several orders of
  magnitude.  Our improved, accurate prefactor and recent advances in
  molecular dynamics simulations and laboratory experiments for argon
  bubble nucleation enable us to precisely constrain the free energy
  barrier for bubble nucleation. Assuming the correction to the CNT
  free energy is of the functional form suggested by Tolman, the
  precise evaluations of the free energy barriers suggest the Tolman
  length is $\simeq 0.3 \sigma$ independently of the temperature for argon
  bubble nucleation, where $\sigma$ is the unit length of the
  Lenard-Jones potential. With this Tolman correction and our new prefactor one
  gets accurate bubble nucleation rate predictions in the parameter range probed
  by current experiments and molecular dynamics simulations.

\end{abstract}

\maketitle

\section{Introduction}

Bubble nucleation in liquid is a liquid-to-vapor transition
\cite{BlanderKatz1975} phenomenon, and plays an important role in many
areas of science and technology, e.g., degassification of steel
\cite{BlanderKatz1975}, bubble jet printers \cite{Iida1994}, vulcanism
\cite{Sparks2008, Toramaru1995,Yamada2005}, the direct detection
of dark matter \cite{Behnke2008, Archambault2011, Felizardo2012,
  Behnke2013} and medicine \cite{Coussios,Blatteau}.
  Bubble nucleation occurs in a metastable liquid under a pressure
  below its equilibrium vapor pressure.  The
liquid phase can even exist at negative pressures, and bubble
nucleation is often observed at negative pressure, although there are
no fundamental differences between a boiling, superheated (positive pressure)
and a cavitating, stretched (negative pressure) liquid \cite{Vinogradov2008}.
Studies of homogeneous liquid-vapour nucleation typically use the
classical nucleation theory (CNT) for the bubble nucleation
rate. However, the range of applicability of the CNT is not well understood.

Numerical techniques such as molecular dynamics and Monte-Carlo
simulations are powerful methods which can resolve details of the nucleation
process and provide useful test cases for nucleation models
\cite{Tanaka2005, Novak2007,Sekine2008,Tsuda2008, Wang2008, 
  Watanabe2010, Tanaka2011, 
 Meadley2012, Abascal2013, Watanabe2013, Diemand2013, Baidakov2014,
  Tanaka2014h, Diemand2014, Angelil2014, Angelil2014h}.
  Typically, these simulations show large
deviations from the CNT predictions. The CNT nucleation rates usually
underestimate the bubble nucleation rates by very large factors \cite{
  Sekine2008,Tsuda2008,Wang2008, Watanabe2010, Meadley2012,
  Abascal2013, Watanabe2013, Baidakov2014,Diemand2014}.  Most of the
simulations for bubble nucleation in the literature use around $10^5$
or fewer atoms, making it difficult to measure nucleation rates
directly.

Recently, Diemand et al.~\cite{Diemand2014} presented large-scale,
micro-canonical molecular dynamics simulations of homogeneous bubble
nucleation with $5 \times 10^8$ Lennard-Jones atoms, and succeeded in
measuring nucleation rates in the range of
$10^{21-25}$cm$^{-3}$s$^{-1}$ for argon by directly resolving bubble nucleation
events in the steady state nucleation phase. The measured rates agree
well with the CNT within two orders of magnitude in the superheated
boiling regime (positive ambient pressure), while the CNT prediction
underestimates the nucleation rates significantly in the 
cavitation regime (lower temperatures and negative pressures).

The kinetics of explosive cavitation in liquid has also been 
investigated in laboratory experiments,
which have measured the superheat temperature for liquid argon at both
positive and negative pressures, by pulse heating liquid around a thin
wire in a negative pressure wave \cite{Vinogradov2008,Vinogradov2009}.
Using this method, Vinogradov et al.~\cite{Vinogradov2008} measured
nucleation rates of $10^{16-18}$cm$^{-3}$s$^{-1}$ in superheated
liquid argon of high purity.

The recent advances in molecular dynamics simulations and experiments
enable us to precisely test theoretical bubble nucleation models and
also to improve them. One of the most serious problems in the CNT model
is that the bulk value of the surface tension is used to evaluate nanobubble formation energy.
Since the
nucleation rate of the CNT depends exponentially on the formation
energy, an incorrect estimate of it can cause a huge error in the
nucleation rate. According to Tolman's correction~\cite{Tolman}, the
surface tension at surfaces of small nuclei (bubbles or droplets) is
dependent on their radius. A model parameter called the Tolman
length can be determined by the measurement of the surface tension of
small nuclei~\cite{Thompson1984, Moody2001,Giessen2002, Lei2005,
 Giessen2009, Sampayo2010, Kuksin2010, Block2010, 
 Horsch2012, Tanaka2014, Diemand2014}. 
 Diemand et al. \cite{Diemand2014} showed that the CNT model agrees well 
 with the nucleation rate from their MD simulations, by using such
a model for the surface tension with a proper Tolman length. The introduction
of a Tolman correction can also significantly improve model predictions in
vapor-to-liquid droplet nucleation~\cite{Tanaka2014}.

In bubble nucleation, additional detail in the treatment of the
process is required, in comparison to droplet nucleation, because of
the variable vapor pressure and density in bubbles as they
grow~\cite{Kagan, Gunther2002,Schmelzer2006,
  Uline2007,Uline2010,Torabi2013}. The vapor pressure in bubbles
varies as they grow and has a significant effect on their growth rates
and on the pre-exponential factor in the CNT formula for the
nucleation rate. Thus we have to solve the two-dimensional evolution
(i.e., the radius and pressure) for the bubble growth process. An
extensive study of this problem has already been done by
Kagan~\cite{Kagan}.  Kagan showed that the pre-exponential factor is
strongly dependent on the ambient liquid pressure reduced from the
saturation. Despite this, a constant prefactor is usually adopted in
the widely-used CNT. Furthermore, the formation energy of a bubble
also depends on the vapor pressure.  Although the vapor pressure is
approximately given by the saturated pressure for critical bubbles, a
more accurate vapor pressure is necessary for the evaluation of the
formation energy in the CNT~\cite{BlanderKatz1975}.

Although these treatments have been developed in individual studies,
they are not included directly in the widely-used CNT. For
comparisons with recent molecular simulations and laboratory
experiments, we should use a precisely crafted expression for the CNT bubble nucleation rate. 
Detailed comparisons also enable us to correctly determine the Tolman length in
the model of the surface tension.

In this paper, we first present a more complete expression for the nucleation 
rate, by summarizing the above studies (Sec. II). Next, we compare the 
improved model with the original CNT or with the measured values in the 
recent MD simulations and experiments (Sec. III). We find that the 
difference between the improved model and the CNT could be several orders 
of magnitude for realistic bubble nucleation parameters.
Comparisons with MD simulations can determine the Tolman length more accurately
thanks to the more accurate pre-factor in our model. 
From our comparisons at various temperatures, 
the Tolman length is obtained as $\simeq 0.3$ $\sigma$,
where $\sigma$ is the unit length of the Lenard-Jones potential.
A summary of our findings can be found in Section IV.

\section{Nucleation theory}

\subsection{Classical expression}

The bubble nucleation rate is the number of stable bubbles formed
per unit time per unit volume and is given by \cite{BlanderKatz1975,Landau}
\begin{eqnarray} 
  J = J_0 n_{\rm e}(i_{\rm c}),
\label{rate2}
\end{eqnarray} 
where $i_{\rm c}$  is the number of gaseous molecule in a critical bubble,
 $J_0$ is the prefactor in the nucleation rate, and
the number density of bubbles $n_{\rm  e}(i)$ is given by 
\begin{eqnarray} 
  n_{\rm e}(i) = n_0 \exp \left( -{ \Delta G(i) \over kT }\right),
\label{equilibrium}
\end{eqnarray} 
where $T$ is the temperature, $k$ is the Boltzmann constant, $\Delta
G(i) $ is the minimum work for the formation of a bubble 
with $i$ molecules and $n_0$ is the number density of liquid
molecules.

In the classical nucleation theory (CNT), 
the minimum work for the formation of the critical bubble 
is 
\begin{eqnarray} 
\Delta G_{\rm CNT} 
 &=&   {16 \pi \gamma^3 \over 3  (P_{\rm eq} - P_{\rm l} )^2 },
\label{deltag-CNT}
\end{eqnarray}
where 
$\gamma$ is the surface tension,  $P_{\rm eq}$ and $P_{\rm l}$ are the
equilibrium vapor pressure at saturation 
 and the liquid pressure, respectively. 
The prefactor $J_0$ is given in the CNT by 
\begin{eqnarray}
J_{\rm 0, CNT}
&=&  \sqrt{ 2 \gamma \over \pi m} ,
\label{prefactor-evap}
\end{eqnarray}
which is the same as in the droplet nucleation.
In the above, $m$ is the molecular mass.
Thus, the widely-used expression of the CNT nucleation rate is 
\begin{eqnarray} 
J_{\rm CNT} 
 &=& \sqrt{2 \gamma \over \pi m} n_0 
\exp \left[ - {16 \pi \gamma^3 \over 3 kT (P_{\rm eq} - P_{\rm l} )^2 } \right].
\label{CNTrate}
\end{eqnarray}
We will describe more accurate expressions for $\Delta G$ and
$J_0$ in subsections B and C.

\subsection{Free energy for bubble formation and the Poynting correction} 

The minimum work $\Delta G $ for the formation of a bubble with radius $r$ 
 is given by \cite{BlanderKatz1975,Landau}
\begin{eqnarray} 
\Delta G =&  {\displaystyle 4 \pi r^3 \over \displaystyle 3} & \left\{ 
{1 \over v_{\rm g}} \left[
\mu_{\rm g}(P_{\rm g}) - \mu_{\rm l}(P_{\rm l}) \right] 
- (P_{\rm g}-P_{\rm l}) \right\}  \nonumber \\
& +&   4  \pi r^2 \gamma ,
\label{exact}
\end{eqnarray} 
$\mu_{\rm g}$ and $\mu_{\rm l}$ are the chemical potentials of the gas
and liquid respectively, $P_{\rm g}$ is the gaseous pressure in the
bubble, and the molar volume of a gas $v_{\rm g}$ is given by
$kT/P_{\rm g}$, assuming the ideal gas. Eq.~(\ref{exact}) also assumes
spherical bubbles. Note that the work for bubble
formation is a function of two variables $(r, P_{\rm g})$ or $(r,
i)$. The number of molecules in the bubble is given by
\begin{eqnarray} 
i ={4 \pi r^3 P_{\rm g}  \over 3kT}. 
\label{molecule-number}
\end{eqnarray} 
Previous studies have investigated the minimum work for a bubble
formation with two variables, taking into account bubble
compressibility\cite{Kagan, Gunther2002,Schmelzer2006,
  Uline2007,Uline2010,Torabi2013}.  Here we use the two variables $r$
and $P_{\rm g}$.

The work $ \Delta G$ has a maximum value along the path of the bubble
growth. Such a maximum point on the path corresponds to the critical
bubble and is given by the saddle point in the two dimensional plane
$(r, P_{\rm g})$ [11]. In the next subsection, we describe the growing
path in the vicinity of the saddle point.  The radius and the internal
pressure of the critical bubble are thus obtained from the conditions:
\begin{equation}
   \left(  \partial \Delta G \over 
 \partial P_{\rm g} \right)_{r} =
 \left(\partial \Delta G \over \partial r \right)_{P_{\rm g}}=0.  
\end{equation}
Since the first-order derivatives of the minimum work are determined by 
\begin{eqnarray} 
 & & \left( \displaystyle \partial \Delta G \over 
\displaystyle \partial r  \right)_{P_{\rm g}} 
 = \nonumber \\
 &&  4 \pi r^2  
 \left[ 
{P_{\rm g} \over kT} [\mu_{\rm g}(P_{\rm g})-\mu_{\rm l}]
- \left(P_{\rm g}-P_{\rm l}- {2 \gamma \over r} \right) \right] 
\label{derivative1}
\end{eqnarray} 
and 
\begin{eqnarray} 
 \left( \partial \Delta G \over \partial P_{\rm g} \right)_{r}  &=&
 {4 \pi r^3 \over 3} [ \mu_{\rm g}(P_{\rm g})-\mu_{\rm l}],
\label{derivative2}
\end{eqnarray} 
the critical size of the bubble $r_{\rm c}$ and the 
critical gas pressure $P_{\rm g,c}$ 
are determined by 
\begin{eqnarray} 
&\mu_{\rm g}(P_{\rm g, c})=\mu_{\rm l},  \label{two-conditions1} \\
&r_c =  \displaystyle \frac{2 \gamma}{P_{\rm g, c}- P_{\rm l}}.
\label{two-conditions2}
\end{eqnarray} 
Eqs. (\ref{two-conditions1}) and (\ref{two-conditions2}) indicate the
chemical equilibrium and the mechanical equilibrium at the saddle
point, respectively.  Using Eqs. (\ref{two-conditions1}) and
(\ref{two-conditions2}), the minimum work for the formation of the
critical bubble is 
\begin{eqnarray} 
\Delta G (r_c , P_{\rm g, c}) = { 4 \pi r_c^2 \gamma \over 3}. 
\label{critical-size}
\end{eqnarray} 

Using Eqs.~(\ref{derivative1})-(\ref{two-conditions2}), 
the second-order derivatives of 
the minimum work at the saddle point are given by 
\begin{eqnarray} 
 \left( \partial^2 \Delta G \over \partial r^2 \right)_{P_{\rm g}}  &=&
-8 \pi \gamma \ \ \  <0, 
\label{second-derivative1}
\end{eqnarray} 
\begin{eqnarray} 
 \left( \partial^2 \Delta G \over \partial P_{\rm g}^2 \right)_{r}  &=&
 {4 \pi r_{\rm c}^3 \over 3 P_{\rm g, c}} \ \ \ >0,  
\label{second-derivative2}
\end{eqnarray} 
and 
\begin{eqnarray} 
\left( \partial^2 \Delta G  \over  \partial r \partial P_{\rm g} \right)
=0. 
\label{second-derivative3}
\end{eqnarray} 
With Eqs.~(\ref{critical-size})-(\ref{second-derivative3}),
 the work for bubble formation around
the critical size is given by~\cite{Debenedetti}
\begin{eqnarray} 
\Delta G  = { 4 \pi r_c^2 \gamma \over 3}
-  4 \pi \gamma (r-r_c)^2 +
 {2 \pi r^3 \over 3 P_{\rm g, c}} (P_{\rm g}-P_{\rm g, c})^2, 
\label{delg-app1}
\end{eqnarray} 
up to second order accuracy in $(r-r_c)$ and $(P_{\rm
  g}-P_{\rm g, c})$.  This expression shows that the point $(r_{\rm
  c}, P_{\rm g,c})$ is indeed the saddle point.
 
The gas pressure, $P_{\rm g,c}$, in the critical bubble is determined by
 Equation (\ref{two-conditions1}).
 Integrating $d \mu =v dP$, we obtain
\begin{eqnarray}
\mu_{\rm g}(P_{\rm g, c})-\mu_{\rm g}(P_{\rm eq})&=& 
kT \ln({P_{\rm g,c} \over P_{\rm eq}}),
\label{ideal-gas} 
\end{eqnarray}
for gas  and 
\begin{eqnarray}
\mu_{\rm l}(P_{\rm l})-\mu_{\rm l}(P_{\rm eq}) &=&
v_{\rm l}(P_{\rm l}-P_{\rm eq}), 
\label{mu_l} 
\end{eqnarray}
for liquid. In Eq.~(\ref{ideal-gas}) 
we used the equation of state for ideal gas, 
$v_{g} = kT/P_{g}$  and $v_l$ is assumed to be constant in Eq.~(\ref{mu_l}).
 Noting $\mu_g (P_{\rm eq})=\mu_l(P_{\rm eq})$,
Eqs.~(\ref{two-conditions1}), (\ref{ideal-gas}) and (\ref{mu_l}) 
yields 
\begin{eqnarray} 
 \ln \left( P_{\rm g, c}  \over P_{\rm eq} \right)= 
{v_{\rm l}  \over v_{\rm eq} } \left( {P_{\rm l}  \over P_{\rm eq}}-1 \right). 
\label{chemical-eq}
\end{eqnarray} 
 In (\ref{chemical-eq}), 
 $v_{\rm eq}(=kT/ P_{\rm eq})$ is the molecular volume of ideal gas at the 
equilibrium pressure.

In the classical theory, the right hand side of
Eq. (\ref{chemical-eq}) is set to be zero because the ratio $v_{\rm
  l}/v_{\rm g}$ is small. This approximation in Eq. (17) gives $P_{\rm
  g,c}=P_{\rm eq}$.  Then, in the CNT, the critical radius is given by
\begin{eqnarray} 
r_{\rm c, CNT} = {2 \gamma \over P_{\rm eq} - P_{\rm l} }.  
\label{critical-CNT2}
\end{eqnarray} 
Blander and Katz \cite{BlanderKatz1975} keep the small term
proportional to $v_{\rm l}/v_{\rm eq}$ in Eq.~(\ref{chemical-eq}) and derive
more accurate expressions for the critical radius and the nucleation
rate than CNT. They termed
this correction to CNT as the \textit{Poynting correction}. 
Here we label this correction as 
\textit{PCNT}.  In PCNT, 
the gaseous pressure in the bubble is obtained as 
\begin{eqnarray} 
 {P_{\rm g, c}  \over  P_{\rm eq}} &=&
 \exp \left( {v_{\rm l} \over v_{\rm eq}}
{ P_{\rm l} -P_{\rm eq} \over P_{\rm eq} } \right) \nonumber \\
\simeq  
1  &+&  { v_{\rm l} \over v_{\rm eq} } \left( \frac{P_{\rm l}- P_{\rm eq}}{P_{\rm eq}} \right)
   + {1 \over 2} \left(
    { v_{\rm l} \over v_{\rm eq} } 
 { P_{\rm l} -P_{\rm eq}  \over  P_{\rm eq} }
\right)^2, 
\label{a4}
\end{eqnarray}
then  we have
\begin{eqnarray} 
  P_{\rm g, c} - P_{\rm l}
=   (P_{\rm eq} - P_{\rm l})  \delta , 
\label{a5}
\end{eqnarray}
where $\delta$ is the Poynting correction factor given by
\begin{eqnarray} 
\delta=\left[ 1- { v_{\rm l} \over v_{\rm eq} }
   +    { P_{\rm eq} -P_{\rm l}  \over  2 P_{\rm eq} }
\left( { v_{\rm l} \over v_{\rm eq} } \right)^2 
\right].  
\label{a6}
\end{eqnarray}
  The expansion in
Eq.~(\ref{a4}) is valid even if $P_{\rm l}$ is far from $P_{\rm eq}$
because of the factor $v_{\rm l} / v_{\rm eq}$ is small $(\sim
0.01-0.1) $.  This indicates that $P_{\rm g} (r_{\rm c})$ is close to
$P_{\rm eq}$ even if $P_{\rm l}$ is far from $P_{\rm eq}$.  But, the
gaseous pressure $P_{\rm g}$ can deviate considerably from $P_{\rm
  eq}(r)$ when $r$ is far from $r_{\rm c}$.  We find that in the Poynting
correction factor derived by Blander and Katz, the second order term
of the small ratio $v_{\rm l}/v_{\rm eq}$ is incorrect.  

Note that the approximation of ideal gas is inaccurate at a high
temperature where the equilibrium pressure $P_{\rm eq}(T)$ is
considerably large. We also derive the the Poynting correction factor
for such a non-ideal case with the first order accuracy. Since $P_{\rm
  g,c}$ is close to $P_{\rm eq}$ for critical bubbles,  $v_{\rm
    gas}$ can be replaced by $v_{\rm eq} (=v_{\rm gas}(P_{\rm eq}))$ in
  the first order approximation and we obtain instead of
  Eq.(\ref{ideal-gas}) 
\begin{eqnarray}
\mu_{\rm g}(P_{\rm g,c})
- \mu_{\rm g}(P_{\rm eq}) = v_{\rm eq} ( P_{\rm g,c} - P_{\rm eq} ). 
\label{non-ideal}
\end{eqnarray}
As for $v_{\rm eq}$, we include the non-ideal effect. Using 
the second virial coefficient $B_2$, the molecular volume $v_{\rm eq}$ 
of non-ideal gas is given by\cite{Landau}
\begin{eqnarray}
v_{\rm eq} = kT/P_{\rm eq} +B_2(T). 
\label{non-ideal2}
\end{eqnarray}
Although we use Eq.~(\ref{non-ideal}) instead of (18), we obtain 
the same $P_{\rm g,c}$ as Eq.~(22) up to  the first order term 
and thus have $\delta = 1-v_{\rm l}/v_{\rm eq}$.
Hence, by using $v_{\rm eq}$ of Eq.~(\ref{non-ideal2}) in Eq.(23), we can obtain
the Poynting correction factor for non-ideal gas 
with first order accuracy.

The critical
radius in the PCNT is obtained as
\begin{eqnarray} 
r_{\rm c,PCNT} = {2 \gamma \over (P_{\rm eq} - P_{\rm l} ) \delta }.
\label{rc-pcnt}
\end{eqnarray}
With this Poynting correction for $r_{\rm c}$,
the nucleation rate is given by 
\begin{eqnarray} 
J =  J_0 n_0 \exp
\left[ - {16 \pi \gamma^3 \over 3 kT (P_{\rm eq} - P_{\rm l} )^2
\delta^2 } \right].
\label{PCNTrate}
\end{eqnarray}
We will consider the prefactor $J_0$ in the next subsection. Although
$1-\delta$ $(\simeq v_{\rm l}/v_{\rm g})$ is usually small, the
difference in the nucleation rate between the CNT and PCNT can be
large because of the strong exponential dependence. This
correction is necessary especially at a relatively high temperature where the
ratio $v_{\rm l}/v_{\rm g}$ is not so small due to high $P_{\rm eq}$.
At such a high saturated pressure, it is also necessary to include 
the non-ideal gas effect for $v_{\rm eq}$ (Eq.~[\ref{non-ideal2}]).

For a more accurate evaluation of the nucleation rate, we
also need to take into account the deviation of the surface tension
from the bulk value for the nano-sized critical bubbles. In Sec. III,
we adopt the Tolman correction for the surface tension [27] and fix
the model parameter, the Tolman length, by using the results of MD
simulations.

\begin{figure}
\includegraphics[height=.33\textheight]{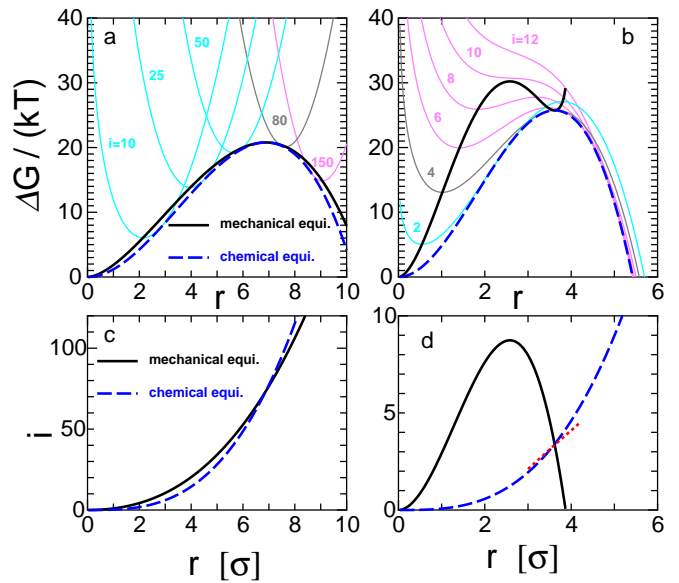}
\caption{(Color online) Free energy for bubble formation $\Delta G$ as
  a function of the bubble radius for Lennard-Jones system with
  $T^*=0.855$ (panel a) and 0.7 (b).  The liquid pressures are
  $P_l=0.017 \varepsilon \sigma^{-3}$ and $P_l=-0.16 \varepsilon
  \sigma^{-3}$ for $T^*=0.855$ and 0.7, respectively (see text for
  detail).  The black solid curves assume the mechanical equilibrium,
  while the blue dashed indicate the chemical equilibrium.  The thin
  solid curves show $\Delta G$ for various constant bubble molecule
  numbers $i$.  Panel c and d show the relations between $i$ and $r$
  in the mechanical and the chemical equilibriums at $T^*=0.855$
  (panel c) and 0.7 (d).  In the positive pressure case with
  $T^*=0.855$, the paths and $\Delta G$s in the two equilibriums are
  close each other, whereas the deviations in them are large in the
  negative pressure case.  Panel d also shows the path of bubble
  growth across the critical size by the doted line.  }\label{deltagr}
\end{figure}

Before proceeding to the evaluation of the prefactor, we show some examples
of $\Delta G$ in Figure~\ref{deltagr}.  For the comparison with
molecular dynamics simulations in Section 3, here we consider a Lennard-Jones liquid.
In Figure~\ref{deltagr}, panels a
and b show $\Delta G$ as a function of the bubble radius $r$ for the
temperatures $T^* ( \equiv kT/\varepsilon) = 0.855$ and 0.7,
respectively, where $\varepsilon$ is the binding energy of the
Lennard-Jones potential.  At each temperature, the liquid pressures
are set to be 0.017 $\varepsilon \sigma^{-3}$ and -0.16 $\varepsilon
\sigma^{-3}$, respectively, where $\sigma$ is the unit length in the
Lennard-Jones potential.  The equilibrium pressures are given by
$0.046 \varepsilon \sigma^{-3}$ and $0.010 \varepsilon \sigma^{-3}$
and the surface tensions are $0.089 \varepsilon \sigma^{-2}$ and $0.33
\varepsilon \sigma^{-2}$ respectively at each
temperature~\cite{Diemand2014}. Here, we set $v_{\rm l}/v_{\rm g} = 0$.
To evaluate $\Delta G$, we also fix $i$ (or $P_{\rm g}$).  In Panels a
and b, we plot $\Delta G$ for bubbles in mechanical equilibrium
(solid lines) and in chemical equilibrium (dashed lines). The thin
solid lines show $\Delta G$ for various constant bubble molecule numbers $i$. Panel c and d
show the relation of $i$ and $r$ for the mechanical and the chemical
equilibriums at $T^* = 0.855$ and 0.7.

In the positive pressure case of Panel a, both $\Delta G$ in the
mechanical and the chemical equilibria are similar to
one another. They agree exactly at their
maxima - corresponding to the critical size. At other radii,
$\Delta G$ is slightly smaller at chemical equilibrium. The
minima of each constant-$i$ line are located on the line
of the mechanical equilibrium. Thus we find that the critical
size corresponds to the saddle point.

In the negative pressure case of Panels b and d, the 
mechanical equilibrium lines deviate considerably
from those in chemical equilibrium. The maximum of $\Delta G$
in the chemical equilibrium corresponds to the minimum of
the mechanical equilibrium case. This also shows that it 
is the saddle point. Panel d shows that the number of
molecules of growing critical bubbles increases for the
chemical equilibrium case whereas it decreases for the
mechanical equilibrium case. These facts indicate that 
chemical equilibrium is more realistic than mechanical
equilibrium. In the next subsection, we examine the growth
of bubbles using the model of Kagan~\cite{Kagan} to obtain
the prefactor. It also gives us the path around the critical
size. In Panel d, we also plot the obtained path across
the critical size (dotted line), which is close to the line of
the chemical equilibrium.

\subsection{The nucleation rate prefactor}

Compared to droplet nucleation, bubble nucleation requires a more
detailed treatment due to the bubble compressibility. The vapor pressure
in the bubbles varies as they grow and significantly affects
their growth rates and the pre-exponential factor in the CNT
expression for the nucleation rate, as well as the bubble formation energy. Kagan~\cite{Kagan} solved the two-dimensional bubble evolution and showed that the pre-exponential factor is strongly
dependent on the ambient liquid pressure.  Based on the method of
Kagan~\cite{Kagan}, we evaluate the exact prefactor $J_0$.

The nucleation rate is usually given by
\begin{eqnarray} 
 J = {1 \over \int [ D n (i)]^{-1} di } \simeq 
 Z D_c n_e(i_c),
\label{rate-exact}
\end{eqnarray} 
where  the Zeldovich factor $Z$ is 
\begin{eqnarray} 
 Z = \left[ - {1 \over 2 \pi kT}
\left( d^2 \Delta G \over di^2 \right)_{i_c}  \right]^{1/2}
\label{zeld}
\end{eqnarray} 
and $D$ is the diffusion coefficient in the $i$-space 
given by~\cite{Kagan} 
\begin{eqnarray} 
 D = -kT {\displaystyle  \left( di \over dt \right)
\Big \slash  \displaystyle \left( d \Delta G \over di \right)}
\label{diffusion}
\end{eqnarray} 
and $D_c=D(i_c)$. Note that $Z$ and $D_{\rm c}$ 
 are evaluated at the critical size $i_{\rm c}$
(or $r_{\rm c}$).  
 Since the denominator
and fraction in Eq.(\ref{diffusion})  both vanish at the critical
size, we evaluate it with the second derivatives
\begin{eqnarray} 
 D_{\rm c} = -kT { \displaystyle {d \over di} \left( di \over dt \right)_{i_c}
 \Big \slash  \displaystyle \left( d^2 \Delta G \over di^2 \right)_{i_c}}.
\label{diffusion2}
\end{eqnarray} 
  
From Eq.~(\ref{rate-exact}), the prefactor in the nucleation rate is
$J_0=Z D_c$.  We use $r$ instead of $i$ in the equations for convenience, and rewrite $J_0$ as
\begin{eqnarray} 
 J_0 = \left(\displaystyle  {kT \over 
 2 \pi} \Big \slash  \displaystyle 
\left| d^2 \Delta G \over dr^2 \right|_{r_c}  \right)^{1/2} A 
\left( di \over dr \right)_{i_c},
\label{prefactor}
\end{eqnarray} 
where 
\begin{eqnarray} 
A= \left[ { d \over dr} \left( dr \over dt \right)\right]_{r_c}. 
\label{A}
\end{eqnarray} 

The growth rate  of a spherical bubble is described by the
Rayleigh-Plesset equation \cite{Angelil2014}, 
\begin{eqnarray}
{r {d^2r \over dt^2}}
= -{3 \over 2 } \left( {dr \over dt} \right)^2
+ { v_{l} \over m } \left(P_{g}-P_{l}-{ 2 \gamma \over r} - {4 \eta \over r}
{dr \over dt} \right), 
\label{RP}
\end{eqnarray}
where $\eta$ is the viscosity of the liquid. 
The evaporation rate from the bubble surface, i.e., the time evolution
of the molecule number in a bubble $i$, is given by~\cite{Kagan,
  BlanderKatz1975}
\begin{eqnarray} 
  {di \over dt} =  4 \pi r^2 { \alpha \over 1 + \epsilon}
{ P_{\rm eq}-P_{\rm g} \over \sqrt{ 2 \pi m kT} }, 
\label{evaporation}
\end{eqnarray} 
where $\alpha$ is the evaporation coefficient (often taken to be
unity).  The factor $\epsilon$ is introduced to include the effect of
the temperature difference between the ambient liquid temperature $T$ and
that at the bubble surface~\cite{Kagan} which is
given by $\epsilon=\alpha d q r_{\rm c} / (\lambda \sqrt{ 2\pi m k T
})$, where $q$ is the latent heat of evaporation per molecule, $\lambda$ is the
thermal conductivity coefficient of the liquid, and $d=d P_{\rm
  eq}/dT$.

Bubbles smaller than the critical size shrink, while larger ones
grow. At the critical size, the growth rate therefore vanishes.
 So we can put
$dr/dt$ around $r \simeq r_{\rm c}$ as
\begin{eqnarray}
{dr  \over dt} = A (r-r_c), 
\label{growthrate}
\end{eqnarray}
Since the vapor pressure in the critical bubble is approximately
equal to $P_{\rm eq}$, the 
 vapor pressure is given by 
\begin{eqnarray}
P_{\rm g}=P_{\rm eq} + {d P_{\rm g}  \over dr} (r-r_c),   
\label{gas-pressure}
\end{eqnarray}
around $r \simeq r_{\rm c}$, 
where we have neglected the small terms proportional to $v_{\rm l}/v_{\rm g}$. 

Substituting Eq.~(\ref{gas-pressure}) into Eq.~(\ref{evaporation}) and
transforming the left hand side of Eq.~(\ref{evaporation}) by the use of
Eqs.(\ref{molecule-number}), (\ref{growthrate}), and
(\ref{gas-pressure}), we obtain
\begin{eqnarray}
{d P_{\rm g} \over dr}
= -{ 1 + \epsilon \over \alpha V_{\rm th}}
 P_{\rm eq} A
\left( 1  + { 1 + \epsilon \over 3 \alpha V_{\rm th}} r_{\rm c} A
\right)^{-1}, 
\label{dpdr}
\end{eqnarray}
where $V_{\rm th}$ is the thermal velocity defined by 
$ \sqrt{kT / (2 \pi m)}$. Substituting
Eqs.~(\ref{growthrate})-(\ref{dpdr}) into Eq.~(\ref{RP}), we obtain the 
equation for $A$:
\begin{eqnarray}
A^2 + {1 \over  \rho_{\rm l} r_{\rm c} } &
 \left [ { \displaystyle 1 + \epsilon \over \displaystyle 
 \alpha V_{\rm th}} P_{\rm eq} 
  \left( 1 +  { \displaystyle 1+\epsilon \over \displaystyle 
 3  \alpha V_{\rm th}} 
  r_{\rm c}  A \right)^{-1} + 
  { \displaystyle 4 \eta  \over \displaystyle 
r_{\rm c}} \right]  A \nonumber \\ & 
 - { \displaystyle 2  \gamma  \over \displaystyle \rho_{\rm l} r_{\rm c}^3} =0.   
\label{RP4}
\end{eqnarray}
Equation~(\ref{RP4}) corresponds to the cubic equation derived by
Kagan~\cite{Kagan}.

Here we present a simple approximate solution to $A$. 
Since the bubble growth rate is significantly slower than the thermal velocity
$V_{\rm th}$, we assume that $A \ll 3 \alpha
V_{\rm th}/ r_{\rm c}/(1+ \epsilon)$.  
Then, the factor in Eq.~(\ref{RP4}) is rewritten as 
\begin{eqnarray}
  \left( 1 +  {  1+\epsilon \over 3  \alpha V_{\rm th}} 
  r_{\rm c}  A \right)^{-1}  \simeq 
1-  { 1+\epsilon \over 3  \alpha V_{\rm th}}  
r_{\rm c}  A 
\end{eqnarray}
which reduces Eq.~(\ref{RP4}) to a
quadratic equation and we obtain an approximate solution
 of $A$ as 
\begin{eqnarray}
A&
=&
 A_{\rm ine}   \nonumber \\ &\times&\left\{
\sqrt{ \left( 
  { A_{\rm ine } \over  2 A_{\rm eva }} +{ A_{\rm ine } \over  2A_{\rm vis }}  
\right)^2 + 1 }
-    { A_{\rm ine } \over  2 A_{\rm eva }} -{ A_{\rm ine } \over  2A_{\rm vis }}  
 \right\}.
\label{prefactor-exact-a}
\end{eqnarray} 
The factors in Eq.~(\ref{prefactor-exact-a}) $A_{\rm eva }$, $A_{\rm vis
}$, and $A_{\rm ine}$ are determined by the evaporation, the inertia
of the fluid motion, and the viscosity at the surface region of the
bubble, respectively, and given by
\begin{eqnarray}
A_{\rm eva } &=& {2 \gamma \over r_{\rm c }^2} 
{\alpha  V_{\rm th } \over (1+ \epsilon)  P_{\rm eq }}, 
\\
A_{\rm vis } &=& { \gamma \over 2 r_{\rm c } \eta}, 
\\
A_{\rm ine } &=& \left( 2 \gamma \over \rho_{\rm l } r_{\rm c }^3 a \right)^{1/2} ,
\label{prefactor-each}
\end{eqnarray}
with
\begin{eqnarray}
a= 1 - {(1+ \epsilon)^2 P_{\rm eq }  \over 3  \alpha^2
 V_{\rm th}^2 \rho_{\rm l}}.  
\label{prefactor-factor}
\end{eqnarray}
In the limit $A_{\rm eva } \ll A_{\rm vis },$ $A_{\rm ine }$, the
growth is regulated by the evaporation process and  $A=A_{\rm evap }$.  If
$A_{\rm ine }$ ($A_{\rm vis }$) is the smallest of them, the growth is
determined by the inertia (the viscosity) of the fluid and $A$ is
given by $A_{\rm ine }$ ($A_{\rm vis }$).  We recall that this
approximate solution of Eq.(\ref{prefactor-exact-a}) is valid in the
case where $A \ll 3 \alpha V_{\rm th }/r_c/(1 + \epsilon)$.

To evaluate $J_0$ of Eq.~(\ref{prefactor}),
 we also need $ d^2 \Delta G/ di^2$  and 
$di/dr$. Using  Eq.~(\ref{dpdr}), we obtain 
\begin{eqnarray}
& \displaystyle \frac{ d^2 \Delta G} {dr^2}
= \left( \partial^2 \Delta G \over  \partial r^2 \right)_{P_{\rm g}} + 
\left( \partial^2 \Delta G \over  \partial P_{\rm g}^2 \right)_{r}
\left( d P_{\rm g} \over  d r  \right)^2 \nonumber \\
& = - 8 \pi \gamma \left[ 1-
{ 3 P_{\rm eq} \over P_{\rm eq}- P_{\rm l}}
\left(1 + { 3 P_{\rm eq} \over P_{\rm eq}- P_{\rm l}} {A_{\rm eva} \over A}
\right)^{-2} \right], 
\end{eqnarray}
and 
\begin{eqnarray}
{ d i \over dr}
=  4 \pi r_{\rm c}^2 
{ P_{\rm eq} \over kT}  \Big \slash \left(1 + 
 {P_{\rm eq}-P_{\rm l} \over 3 P_{\rm eq} } {A \over A_{\rm eva}}
\right).
\label{didr}
\end{eqnarray}
Their $P_{\rm eq}$-dependencies affect $J_0$. 
Thus by the use of $A$, we find the expression for the prefactor in the
 nucleation rate: 
\begin{eqnarray}
J_0&=& { \alpha \over 1+ \epsilon} \sqrt{ 2 \gamma \over \pi m}
{ { \displaystyle A \over \displaystyle A_{\rm eva}} 
 \over 1 + {\displaystyle P_{\rm eq}- P_{\rm l} \over \displaystyle 3P_{\rm eq}}
 {\displaystyle A \over \displaystyle A_{\rm eva}} }\nonumber \\
& \times & \left[ 1 - {3 P_{\rm eq} \over P_{\rm eq} - P_{\rm l}}
   \left( 1 + { 3 P_{\rm eq} \over P_{\rm eq} - P_{\rm l}}
 {A_{\rm eva} \over A }  \right)^{-2} \right]^{-1/2}_.  
\label{prefactor-exact}
\end{eqnarray}
Note that the explicit expression of the exact prefactor 
 was not presented by Kagan~\cite{Kagan}. 

Fig. 2 shows the prefactor given by Eq.~(\ref{prefactor-exact}) as a
function of the liquid pressure $P_{\rm l}$ for Lennard-Jones liquid
with $T^*=0.6, 0.7,$ and 0.855.  The equilibrium pressure and the
surface energy are given by $ P_{\rm eq}=0.0034 \varepsilon
\sigma^{-3}$ and $\gamma=0.51 \varepsilon \sigma^{-2}$, respectively
at $T^* = 0.6$~\cite{Diemand2014}. Here, the evaporation coefficient $\alpha$ is set to unity and $\epsilon$ to zero.  
The viscosity $\eta$ is set to be 0.6 $\varepsilon \tau \sigma^{-3}$ 
with the time unit $\tau$ 
according to Ang\'elil et al.~\cite{Angelil2014}.

When the liquid pressure is very
close to the equilibrium value, $J_0$ agrees with the CNT value of
Eq.~(\ref{prefactor-evap}).  In this case, the bubble growth is
regulated by the evaporation, and the mechanical equilibrium is almost
satisfied on the path of bubble growth near the critical size.  For a
large negative pressure, on the other hand, the prefactor can be significantly
smaller than the CNT value by a factor $ \sim 10^{3}$ because of the
viscosity and inertia effects in the liquid. In this case, chemical
equilibrium holds rather than the mechanical one on the growing path.
We obtain this growing path near the critical size in the $r-i$ plane,
by evaluating $di/dr$ with Eq.~(\ref{didr}).  The obtained path is
plotted in Fig.~1d.

\begin{figure}
\includegraphics[height=.55\textheight]{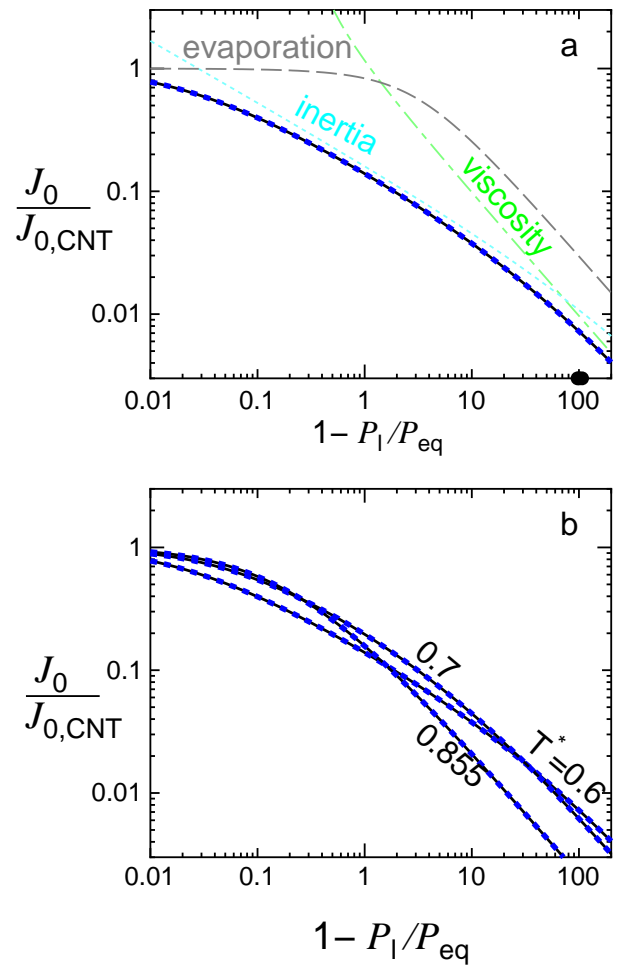}
\caption{(Color online) The prefactor obtained by
  Eq.~(\ref{prefactor-exact}) as a function of the liquid pressure
  $P_{\rm l}$.  Panel a shows the case of $T^*=0.6$ (see text for
  detail).  The solid line presents the prefactor with the exact
  solution of $A$ to Eq.~(\ref{RP4}) whereas the dashed line uses the
  approximate solution of Eq.~(\ref{prefactor-exact-a}).
  Eq.~(\ref{prefactor-exact-a}) reproduced very well the exact
  solution within the accuracy of 0.1 \%.  The prefactors calculated
  with $A=A_{\rm eva}, A_{\rm ine},$ or $A_{\rm vis}$ are also plotted
  with gray dashed, light blue dotted, 
 or green dotted-dashed curves.  The liquid pressures
  used in the molecular dynamics simulations (see Sec III) are plotted
  by circles on the horizontal axis.  The prefactors for various
  temperatures $T^*=0.6, 0.7$ and 0.855 are shown in panel b.
}\label{prefactor-fig}
\end{figure}

\section{Comparison between the classical nucleation rate and the
exact formula }

\begin{figure}
\includegraphics[height=.4\textheight]{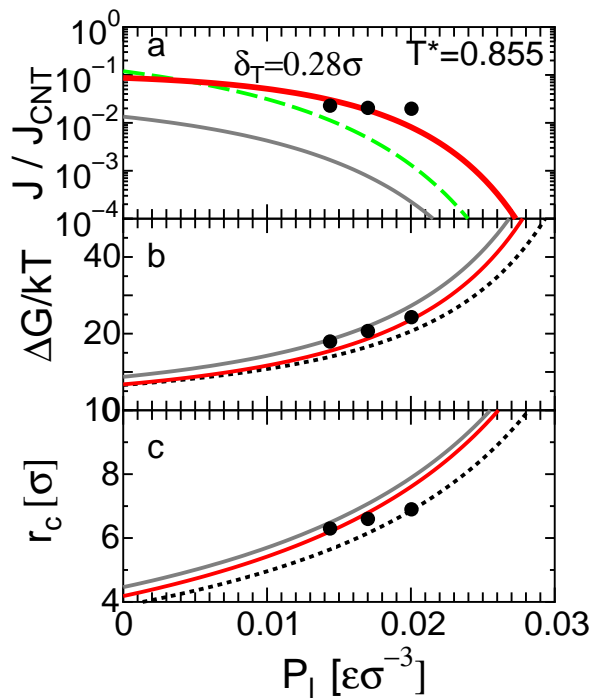}
\caption{(Color online) Comparison between the improved formula and
  the CNT for (TSF) Lennard-Jones fluid at $T^*=0.855$. Panel a shows
  the comparison in $J$.  The improved nucleation rate given by
  Eqs.~(\ref{PCNTrate}) and (\ref{prefactor-exact}) without the Tolman
  correction is plotted with the gray line.  The
  improved formula with the Tolman correction is the thick red curve,
  where $\delta_T$ is set to be $0.28\sigma$.  The nucleation rate
  given by Eq.(\ref{PCNTrate}) with $J_0=J_{\rm 0,CNT}$ is the green
  dashed curve.  Panel b shows the free energy for formation of a
  critical bubble for the PCNT without and with the Tolman correction
  (gray and red solid curves) and for the CNT (dotted
  curve).  Panel c shows the critical bubble radii
 (same models and line-styles as above).
 In all panels, the results obtained by the molecular
  dynamics simulations~\cite{Diemand2014} are plotted with filled
  circles.  }\label{comp0.85}
\end{figure}

We first compare the theoretical nucleation models described in Sec II
with the molecular dynamics simulations by \cite{Diemand2014}.  The MD
simulations used molecules with a truncated force-shifted (TSF)
Lennard-Jones potential and a cutoff length of $2.5 \sigma$. In the
models, we thus use the values of TSF Lennard-Jones liquid for the
thermodynamics data such as the surface tension and the equilibrium
pressure. The evaporation coefficient $\alpha$ is assumed to be unity.
In the calculations of the improved nucleation rate, we use
Eq.(\ref{PCNTrate}).  In Figs.~\ref{comp0.85}, we show comparisons at
$T^* = 0.855$. We plot the results of MD simulations (with filled
circles) and four theoretical models.  The first model is the CNT. The
CNT nucleation rate is given by Eq.~(\ref{CNTrate}).  Other nucleation
rates are normalized with respect to this in Fig.~\ref{comp0.85}a.
The second (green dashed curve) is the PCNT nucleation rate which is
given by Eq. (\ref{PCNTrate}) with $J_0 = J_{\rm 0, CNT}$ of
Eq.~(\ref{prefactor-evap}).  The Poynting correction factor is
evaluated from Eqs.~(\ref{a6}) and (\ref{non-ideal2}) to include the
non-ideal gas effect.

The third (gray curve) is our improved expression given by
Eqs. (\ref{PCNTrate}) and (\ref{prefactor-exact}).  The last one (red
curve) uses the same improved expression as the third one, but also
includes the Tolman correction to the surface tension.  The Tolman
correction describes the size-dependent surface tension as
\begin{equation}
\gamma=\gamma_{\infty}/(1+2 \delta_T /r),
\label{tolman-surface} 
\end{equation}
where $\gamma_{\infty}$ is the surface energy of the planar interface.
The Tolman length in the Tolman correction $\delta_T$ expresses the
curvature dependence on the surface tension of the bubble.  If
$\delta_T$ is positive, the surface tension of small bubbles or
droplets is smaller than the planar one.

In Fig.~\ref{comp0.85}a, we find that the improved formula (gray
curve) can be smaller than the CNT (the PCNT) by a factor of $ \sim
10^{-4}$ ($\sim 1/5$).  Fig.~\ref{comp0.85} shows that the Tolman
correction is necessary to reproduce the results of the MD
simulations.  From the fitting, we obtain $\delta_T = 0.28\sigma$.
Fig.~\ref{comp0.85}b shows the peak values of the free energy for
bubble formation. We find the PCNT correction almost reproduces
$\Delta G(r_{\rm c})$ in the MD simulations.  
  Fig.~\ref{comp0.85}c shows the critical bubbles, $r_c$, obtained
  from the theoretical models and MD simulations.  All models
  successfully reproduce the $r_c$ measured in the MD simulations.
  Note that the Tolman length of $0.28 \sigma$ obtained by fitting
  is much smaller the critical radius.  Thus the correction to the
  surface tension of Eq.~(\ref{tolman-surface}) contributes only at
  the 10~\% level, yet significantly affects the nucleation rate due
  to the high sensitivity that the surface tension has on $J$ (see
  Eq.~(\ref{PCNTrate})).

\begin{figure}
\includegraphics[height=.4\textheight]{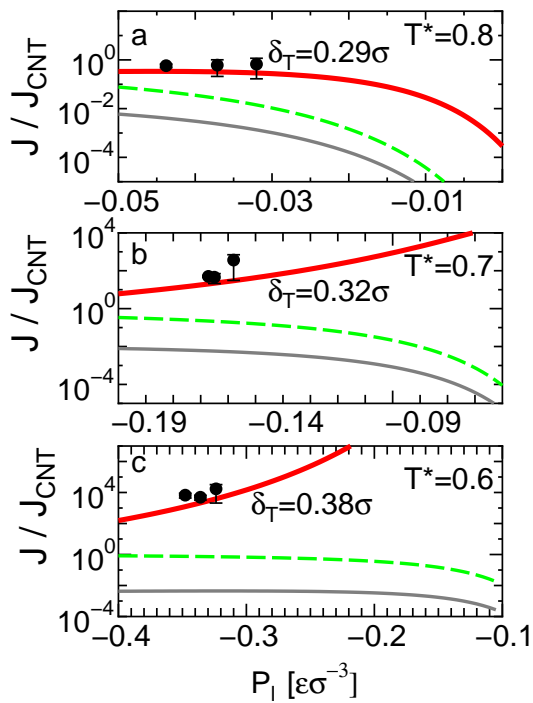} 
\caption{(Color online) The same as Fig.\ref{comp0.85}(a), but for
  $T^*=0.8$, 0.7, and 0.6.  In the improved formula with the Tolman
  correction (red curves), the Tolman length $\delta_T$ are set to be
  $\delta_T=0.29, 0.32$, and 0.38 $\sigma$ for $T^*=0.8$, 0.7, and 0.6,
  respectively.  }\label{comp0.7}
\end{figure}

Fig.~\ref{comp0.7} shows the comparisons in $J$ at $T^*=0.6, 0.7,$ and
0.8, where the equilibrium pressure and the surface energy are given
by $ P_{\rm eq}=0.0303 \varepsilon \sigma^{-3}$ and $\gamma=0.17
\varepsilon \sigma^{-2}$, respectively at $T^* =
0.8$~\cite{Diemand2014}.  We find that deviations in the improved
formula (gray curves) from the CNT can be several orders of magnitude
in all cases.  Compared with the values obtained by MD simulations,
the improved formula without the Tolman correction gives significantly
lower nucleation rates.  By the fitting, we obtain $\delta_T = 0.29
\sigma, 0.32\sigma,$ and $0.38 \sigma$ for $T^* = 0.8, 0.7$, and 0.6,
respectively.  The fits at $T^* =0.6-0.855$ show that the Tolman
length depends weakly on temperature.  Since the radii of the critical
bubbles are 3-7~$\sigma$ in the MD simulations, we see from
Eq.~(\ref{tolman-surface}) that the obtained Tolman lengths of $\simeq
0.3 \sigma$ correspond to corrections of 10-20 \% to the surface
tension from the bulk values. As seen in Figs.~\ref{comp0.85} and
\ref{comp0.7}, these small corrections to the surface tension actually
improve much the predictions of the nucleation-rate formula at 
$T^*=0.6-0.855$.

\begin{figure}
\includegraphics[height=.3\textheight]{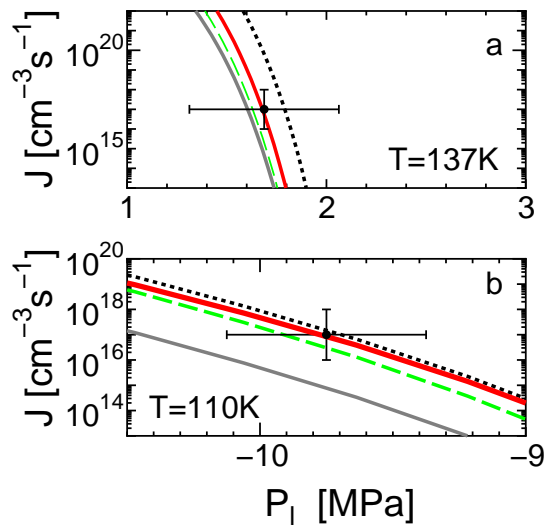}
\caption{(Color online) Comparison between the theoretical models and
  laboratory experiments in $J$ for argon at $T = 137$~K (a) and 110 K
  (b).  The improved nucleation rates without the Tolman correction and 
with the Tolman correction of $\delta_T = 0.3 \sigma$ at $T = 137$~K
 ($\delta_T = 0.13 \sigma$ at $T = 110$~K) are
shown by the gray and red solid curves, respectively. 
  The nucleation rate given by Eq.(\ref{PCNTrate}) with
  $J_0=J_{\rm 0,evap}$ is shown by the green dashed curve. 
  The CNT is shown by the dotted curve.  In the thermodynamics data
  such as the surface tension and the equilibrium pressure, we use the
  data of argon \cite{Iland2007}.  We also show the experimental
  results \cite{Vinogradov2008} (circle), where the error in the
  liquid pressure is determined from the two points at $T=137$~K in
  \cite{Vinogradov2008}.  }\label{exp}
\end{figure}

We also make a comparison with the argon bubble nucleation experiments
by \cite{Vinogradov2008}.  Fig.~\ref{exp} shows the comparisons in $J$
at $T = 137$~K and 110~K.  As for the thermodynamics data in models
such as the surface tension and the equilibrium pressure, we use the
data of argon~\cite{Iland2007}.  We find the deviations between the CNT
and the improved formula can be several order of magnitudes in these
cases, too.

The experimental result is consistent with all of the theoretical
predictions within the error at $T=137$~K.  Thus it is difficult to
fix the Tolman length in this case.  For $T = 110$~K, the improved
formula gives significantly smaller nucleation rates than the
experimental result.  The fitting with the Tolman correction indicates
that the Tolman length $\delta_T$ is $0.13 \sigma \pm 0.14 \sigma$. The error
in $\delta_T$ comes from the errors in $J$ and $P_l$ in the laboratory
experiment.  This value of the Tolman length can be marginally 
consistent with the ones from the fits to the MD simulations.
To further constrain $\delta_T$ more experimental data are needed.
\\

In Fig. 6, we plot the Tolman length for bubbles obtained 
from our analysis (red circles) as a function of $T^*$ as well 
as those in the previous studies (other red symbols).
As for our fitting data with the argon experiment in Fig. 5b 
(red open circle), the normalized temperature $T^*$ is obtained 
as 0.92, because $\epsilon/k = 120$~K for argon. Our results 
indicate that the Tolman length is almost constant
with temperature. Our results agree quite well
with the previous results by  \cite{Kuksin2010}.

Baidakov and Bobrov \cite{Baidakov2014} and 
  Block\cite{Block2010}  obtained small negative values 
 in their MD and MC  simulations, respectively.  
The deviation in the Tolman length between their results and 
ours comes from the different definitions of $\delta_T$
(or the different curvature dependence of the surface tension).
 Their obtained surface tensions are always less than the bulk values, 
 which is consistent with our curvature dependence with a positive
Tolman length. For example, 
Baidakov and Bobrov \cite{Baidakov2014}
 gave the curvature dependence of the 
surface tension as
\begin{equation}
\gamma = \gamma_\infty / (1 + 2\delta_T/r + l^2/r^2).
\label{tolman-surface-2} 
\end{equation}
If Eq.~(\ref{tolman-surface}) is used for the evaluation of $\delta_T$
instead of their curvature dependence,  their results give
 positive $\delta_T$. 
 In fact Eq.~(\ref{tolman-surface}) with $\delta_T =0.3
\sigma$ also very successfully reproduces the surface tension
measurements  by Baidakov and Bobrov \cite{Baidakov2014} 
(in their figure 9).  This means that the results by
Baidakov and Bobrov also agree well with ours. 

  Moody and Attard~\cite{Moody2001} also obtained
 negative Tolman lengths at  $T^*
  \ge 1$ from their MC simulations.  Even at the high
temperatures, nevertheless, their obtained surface tension increases
with the bubble radius at $r \siml 2\sigma$ (see their Figure 10 and
11), which is also consistent with our curvature dependence.

Recent calculations with density functional theory 
\cite{Blokhuis2013, Wilhelmsen2015}
also result in negative Tolman lengths, by using
a different curvature dependence similar to~\cite{Baidakov2014}. 
However, they also obtained smaller surface tensions
than the bulk and, in this sense, their results are 
consistent with ours.

\begin{figure}
\includegraphics[height=.3\textheight]{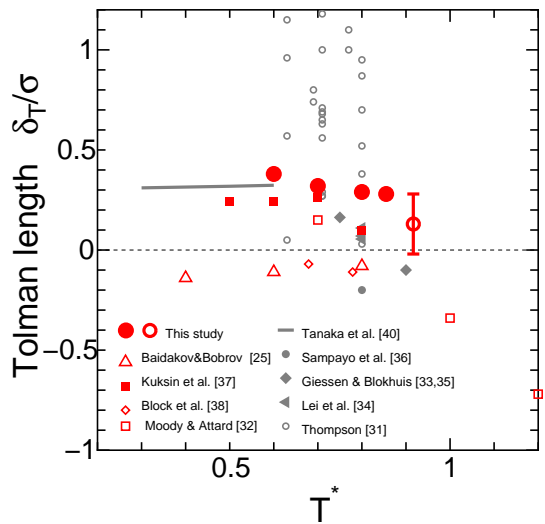}
\caption{(Color online) The Tolman length for bubbles as a function of
  temperature obtained by this study and the previous
  ones~\cite{Moody2001,Kuksin2010,Block2010,Baidakov2014} (red symbols).  Our
  analysis with MD simulations (filled circles) and the experiment at
  110 K (open circle) suggest the Tolman length is approximately given by
 $\delta_T=0.3$~$\sigma$. 
  For reference, the Tolman length for droplets are also shown with gray 
  symbols \cite{Thompson1984, Giessen2002, Lei2005, Giessen2009,
 Sampayo2010, Tanaka2014}.
 }\label{tolman-length}
\end{figure}

For reference, we plot the Tolman length for droplets in
Fig.~\ref{tolman-length}.  We
find the value for droplets by \cite{Tanaka2014} is consistent
with our results for bubbles, although there is some scatter in
the previous values.
Our results suggest that the Tolman length is approximately 
given by $ \simeq 0.3 \sigma$, and is temperature independent.
The Tolman correction with $\delta_T=0.3$~$\sigma$ significantly improves 
the prediction of the nucleation rate with our nucleation rate model.

\section{Conclusion}

We have revised the expression of the bubble
nucleation rate based on the classical theory. In  
bubble nucleation, the prefactor is far more complex than 
in droplet nucleation because bubble growth is 
regulated by many processes (i.e., evaporation, thermal 
conduction, viscosity, and inertia of liquid). This difference
in the prefactor between the bubble and droplet cases has been 
overlooked in many studies. We have also compared the improved 
expression of the nucleation rate with results of the MD 
simulations and laboratory experiments. Our findings are 
summarized below.

\begin{itemize}

\item
In bubble nucleation, the prefactor is strongly 
dependent on the degree of non-equilibrium, i.e., 
the liquid pressure, whereas it is constant in the droplet case.
In the case of a large negative liquid pressure (or the highly 
viscous case), the prefactor can be far smaller than the 
droplet case by a factor of $10^{-3}-10^{-1}$ (Fig. 2).

\item
  When the liquid pressure is slightly below the equilibrium pressure,
  the deviations in the free energy for bubble formation from the CNT
  become large.  For example, in dark matter detection experiments
  which use superheated liquids as targets
  \cite{Behnke2008,Archambault2011,Felizardo2012,Behnke2013}, the
  degree of non-equilibrium is very small. For such a 
  near-equilibrium case, the PCNT should be used instead of CNT.

\item
Comparisons of our improved expression for the nucleation rates $J$ to results
from MD simulations and laboratory experiments suggest that 
the surface tension depends on the bubble size at the nano-scale level. 
The improved expression including the Tolman correction
to the surface tension with a small Tolman length
of $\simeq 0.3 \sigma$ leads to good agreements with the
recent MD simulations and laboratory experiments (Fig. 6). 

\end{itemize}

\begin{acknowledgments}

  We thank two anonymous reviewers for valuable suggestions.  This
  work was supported in part by JSPS KAKENHI Grant Number 26108503,
  2540054, and 26287101.  
  J.D. and R.A. are supported by the Swiss National Science Foundation (SNF).\\
\end{acknowledgments}

\appendix




\bibliography{jcp.bib}   
\bibliographystyle{aipprocl} 


\end{document}